\documentclass[conference]{IEEEtran}
\IEEEoverridecommandlockouts 

\usepackage{booktabs} 
\usepackage{tabularx} 
\usepackage{balance} 
\usepackage{multirow} 
\usepackage{siunitx} 
\usepackage{hyperref} 
\usepackage{graphicx,cleveref}
\newcommand{\multicell}[2][t]{\begin{tabular}[#1]{@{}l@{}}#2\end{tabular}} 
\newcommand{\ra}[1]{\renewcommand{\arraystretch}{#1}} 
\usepackage{tikz}
\newcommand\copyrighttext{%
  \footnotesize
  C. Matthies, R. Teusner and G. Hesse, ``Beyond Surveys: Analyzing Software Development Artifacts to Assess Teaching Efforts,'' 2018 IEEE Frontiers in Education Conference (FIE), San Jose, CA, USA, 2018, pp. 1-9.
  doi: \href{https://doi.org/10.1109/FIE.2018.8659205}{10.1109/FIE.2018.8659205}.
  IEEE Xplore: \url{https://ieeexplore.ieee.org/document/8659205}.
  
  \vspace{0.5em}
  
  Copyright \textcopyright 2018 IEEE. Personal use of this material is permitted.
  Permission from IEEE must be obtained for all other uses, in any current or future  media, including reprinting/republishing this material for advertising or promotional purposes, creating new collective works, for resale or redistribution to servers or lists, or reuse of any copyrighted component of this work in other works.
  }
\newcommand\copyrightnotice{%
\begin{tikzpicture}[remember picture,overlay]
\node[anchor=south,yshift=10pt] at (current page.south) {\fbox{\parbox{\dimexpr\textwidth-\fboxsep-\fboxrule\relax}{\copyrighttext}}};
\end{tikzpicture}%
}

\begin{document}

\title{Beyond Surveys: Analyzing Software Development Artifacts to Assess Teaching Efforts}

\author{\IEEEauthorblockN{Christoph Matthies, Ralf Teusner, Guenter Hesse}
\IEEEauthorblockA{Hasso Plattner Institute \\
University of Potsdam, Germany\\
\{firstname.lastname\}@hpi.de}
}

\maketitle
\copyrightnotice
\begin{abstract}
This Innovative Practice Full Paper presents an approach of using software development artifacts to gauge student behavior and the effectiveness of changes to curriculum design.
There is an ongoing need to adapt university courses to changing requirements and shifts in industry.
As an educator it is therefore vital to have access to methods, with which to ascertain the effects of curriculum design changes.
In this paper, we present our approach of analyzing software repositories in order to gauge student behavior during project work.
We evaluate this approach in a case study of a university undergraduate software development course teaching agile development methodologies.
Surveys revealed positive attitudes towards the course and the change of employed development methodology from Scrum to Kanban.
However, surveys were not usable to ascertain the degree to which students had adapted their workflows and whether they had done so in accordance with course goals.
Therefore, we analyzed students' software repository data, which represents information that can be collected by educators to reveal insights into learning successes and detailed student behavior.
We analyze the software repositories created during the last five courses, and evaluate differences in workflows between Kanban and Scrum usage.
\end{abstract}


\IEEEpeerreviewmaketitle

\section{Introduction}
One of the main goals of universities is to provide students with an education of the best quality possible.
As such, there is the constant need to improve the learning experience in courses, update course contents to changing requirements and strive for more effective organizational structures~\cite{Felder2009}.
With the continuing rise of digitization in universities, an ever-expanding amount of data on learners is available~\cite{Dutt2017, citeulike:3169799}.
Access to new data sources has led to drastic changes both in science and business.
Equally, learning scientists can greatly benefit from having large repositories of educational data available~\cite{citeulike:3169799}.
Analysis of these repositories in order to tackle educational research issues has given rise to the field of Educational Data Mining (EDM)~\cite{Romero2010}.
In this paper, we show how techniques from this domain can be used to gain insights into students' workflows and ascertain whether changes in curriculum design had the desired effects, without relying on traditional surveys alone.

\subsection{Educational Data Mining}
The International Educational Data Mining Society defines EDM as a discipline concerned with ``developing methods for exploring the unique and increasingly large-scale data that come from educational settings and using those methods to better understand students, and the settings which they learn in''~\cite{InternationalEducationalDataMiningSociety2018}.
Traditional student assessment and evaluation methods such as standardized exams can only provide information on specific student traits at certain points in time.
In order to obtain information that could explain students' progress continued recording of their activities using more sophisticated techniques is required.
If designed well, such measurements can provide insights into how students behave, communicate, and participate in learning activities~\cite{Ashenafi2015}.
In an educational setup, data is generated using a variety of disparate systems~\cite{Dutt2017}.
Some settings in which EDM has been applied include:
\begin{itemize}
    \item E-learning and learning management systems, e.g. Moodle~\cite{Gamulin2013}, where web mining approaches have been applied to student data in log files and databases~\cite{Romero2010}.
    \item Massive Open Online Courses (MOOCs), where student collaborations and interactions with the platforms are studied~\cite{Teusner2017a}.
    \item Offline education, where students are lectured in a traditional face-to-face manner. Statistical analyses are applied to students' data, like test scores or peer assessments, which are gathered in classroom environments~\cite{Romero2010}.
\end{itemize}
Analyzing this educational data can help evaluate, validate and eventually improve courses and educational systems, paving the way for a more effective learning process~\cite{Romero2004}.

\subsection{Educational Data Sources}
Educational contexts such as e-learning or MOOCs benefit from student data that is easily available for analysis and can be used to improve courses~\cite{Dutt2017}. 
In these settings, student activity takes place in controlled, digitized setups that data can be extracted from by logging interactions~\cite{Ashenafi2015,Feng2005}.
In in-person university courses or other offline educational settings other data collection strategies have to be employed, for example, the results of peer assessments~\cite{Topping2010} or surveys~\cite{Felder2009}.
In these analog settings, the lack of detailed, high-resolution data on learners can be compensated by collecting additional data sources specific to the individual context, e.g. social network data~\cite{Fire2012} or even the complexity of teacher's lecture notes~\cite{Keen2009}.
However, due to the limited classroom time available, teachers are often forced to choose between spending time assisting students and spending time assessing students and collecting data~\cite{Feng2005}.
In order to alleviate this problem, data that is already being created by students during project work, especially if digital systems are used, can be analyzed in depth.
While this type of already existent data has previously been used to assess students~\cite{Johnson04,Matthies2016a,zazworka2010developers}, it is also valuable to assess the effectiveness of changes in curriculum design and in order to improve classroom courses.

\subsection{Case Study}
In this paper, we describe a case study on how software development artifacts, created during students' project work in a software engineering course, can be used to check educator's assumptions on student behavior.
The course's setting of collaborative software engineering in a simulated real-world scenario is ideally suited for collecting development data.
During the project work, students use common development tools such as version control systems (VCS), issue trackers and Continuous Integration services.
The artifacts produced using these systems, i.e. commits in a VCS containing code changes and descriptions, contain a large amount of information on how students work and collaborate in their groups~\cite{Rosen:2015:FSE,Santos2016}.

\subsection{Research Questions}
The following research questions (RQ) guide our work:
\begin{enumerate}
\renewcommand{\labelenumi}{\textbf{RQ\arabic{enumi}}}
    \item \label{r1} How can surveys be used to gauge students perceptions of changes in course design over time?
    \item \label{r2} What data can be collected from software development artifacts created by students during a classroom course?
    \item \label{r3} What metrics can be applied to student development data to gauge changes in student behavior during project work?
\end{enumerate}

The rest of the paper is structured as follows:
Section~\ref{sec:case_study_context} introduces the university course in which the case study was performed and describes the software development process that was followed. 
Section~\ref{sec:surveys} presents the surveys that were conducted and discusses the perceptions and attitudes of students.
The following Section~\ref{sec:artifact} describes the analysis of students' software development artifacts produced in the course installments of the last five years and discusses the results.
Section~\ref{sec:related_work} presents related work in the field of studies in student behaviors and sources of educational data.
Section~\ref{sec:conclusion} concludes and summarizes our findings.

\section{Case Study Context}
\label{sec:case_study_context}
The undergraduate software development course described in this case study has been running in our university for more than 5 years.
It is repeatedly run in the winter semester with a length of 15 weeks and was most recently taught in the winter semester of 2017/18.

\subsection{Software Engineering Course}
The main goal of the capstone course ``Software Engineering II'' is teaching iterative, agile development methods and best practices in a hands-on fashion, which has become standard practice in universities~\cite{Paasivaara2017,Dzvonyar2018}.
Each week of the course students are expected to work 8 hours on the project including lectures and team meetings.
In a simulated real-world scenario, students are encouraged to apply the agile processes, introduced in lectures and exercises, and adapt them to suit their teams.
The main learning targets of the course include:
\begin{itemize}
    \item Gaining experience with the artifacts and meetings of agile methods
    \item Acquiring knowledge of source code management (SCM) and continuous integration (CI) systems
    \item Developing critical self-assessment skills regarding students own roles in a software development team
\end{itemize}

All participants, who form their own development teams, jointly develop a single software system.
This means students need to communicate and collaborate within their teams as well as with other student teams.
The project is hosted on the public collaboration platform GitHub\footnote{\url{https://github.com/}}, allowing all stakeholders access to the code and documentation.
Junior research assistants, acting as tutors, are present during team meetings and provide advice and assistance.
Regular lectures on agile methodologies and their applications as well as more general software development topics, such as Continuous Integration and testing, take place during the course.

\subsection{Course Evolution}
Course installments prior to the winter semester of 2014/15 taught exclusively the Scrum methodology~\cite{Schwaber2017}.
However, Lean development approaches, such as Kanban~\cite{james1991machine}, have gained popularity in industry~\cite{stateofagile11,Komus2017}.
Therefore, in an ongoing effort to keep the course as relevant and closely related to real-world scenarios as possible, the practice of Kanban was included.
Students employed the Scrum methodology at the beginning of the course, before switching to Kanban.
In comparison to Scrum, Kanban is considered less authoritative and prescriptive, having fewer rituals and rules than Scrum~\cite{Kniberg2009}.
In order to improve learning results, it is therefore advisable to introduce Kanban after students have already gained experience with the more structured Scrum method~\cite{Mahnic2015a}.
Iterations of the course prior to the winter term 2015/16 did not include Kanban and focused solely on the application of Scrum.
The inclusion of Kanban in the course is a major change as it impacts how students collaborate, plan their work and organize their team structures and meetings.
It needs to be evaluated in order to gauge how effective the introduction of a different software development methodology was, both in terms of student satisfaction as well as how well the method was applied in the project.

\subsection{Switching Development Processes}
\begin{figure}[htb]
\centering
    \includegraphics[width=\columnwidth]{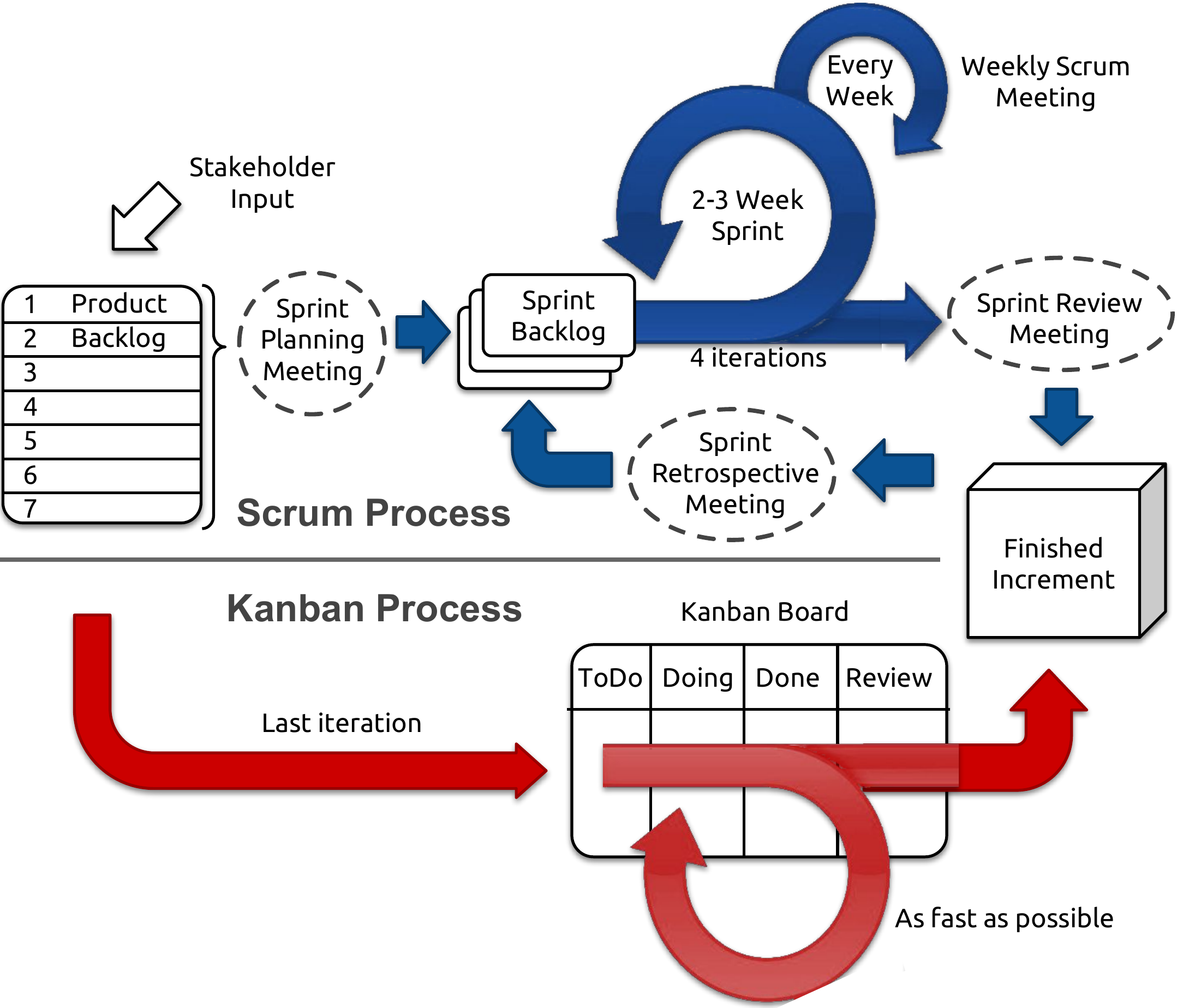}
    \caption{Overview of the modified Scrum process (top, blue) and the Kanban process (bottom, red) used in our software engineering course.}
    \label{fig:process}
\end{figure}

In the first four development iterations of the course, the majority of the course, a modified version of the Scrum process, adapted for the limited time allotted to students for the course is employed.
It is depicted in Figure~\ref{fig:process} at the top.

Participants form self-organizing teams~\cite{Hoda2010} of up to 8 members.
For every Scrum team the roles of Product Owner (PO) and a Scrum Master (SM),  are also performed by student team members, while all other students act as software developers.
In every development iteration (i.e. Sprint), a planning, sprint review, retrospective as well as a weekly synchronization meeting, a stand-up meeting, is organized by the teams.
After course participants have become familiar with the Scrum process and their teams, i.e. after they have reached the \emph{norming} stage of group development~\cite{Bonebright2010}, and have developed a cohesive group, Kanban and its practices are introduced in a lecture.
The concepts of Kanban such as the Kanban board, the idea of workflow visualization and the guiding principle of limiting work in progress (WIP)~\cite{Ahmad2013} are introduced.
We encourage students to try out and apply these new ideas in their teams.
Participants employ Kanban for the last iteration of the project, instead of a final Scrum sprint, see the bottom of Figure~\ref{fig:process}.

\section{Surveys}
\label{sec:surveys}
When trying to assess the impact of changes in curriculum design and whether the expected changes to student behavior took place, students' perceptions can be collected through the use of surveys.

\subsection{End-of-term Survey}
As part of an ongoing effort to collect feedback from students to improve teaching and university courses, standardized end-of-term surveys were conducted in all iterations of our software engineering courses in the years 2013 up to 2018.
This has become standard practice for educational institutions to evaluate teaching quality~\cite{Felder2009}.
The survey is administered online before students receive their final course grades, in order to prevent interference.
The survey collects perceptions of students on a range of topics, including satisfaction with the course in general, perceived importance of course contents and satisfaction with mentoring.
An extract of the questions relevant to student satisfaction with the project work and the course over the iterations of the course is shown in Table~\ref{table:evap_questions}.

\begin{table}[htb]
    \ra{1.2}
    \centering
    \begin{tabular}{@{}ll@{}}
        \toprule
        \textbf{\#} & \textbf{Survey item}\\
        \midrule
        1 & The course was fun\\
        2 & The course motivated me to delve deeper into the discussed topics\\
        3 & I learned a lot in the course\\
        4 & The course is important to my course of studies\\
        5 & The course was well structured\\ 
        6 & The topics of the course were well chosen\\
        7 & How would you rate the course overall?\\
        \bottomrule
    \end{tabular}
    \vspace{0.1cm}
    \caption{Questions and statements of the end-of-term survey.}
    \label{table:evap_questions}
\end{table}

Questions and statements could be rated on a scale of ``fully agree''/``great'' to ``totally disagree''/``bad''.
Results of the anonymous survey are presented to course instructors in aggregate form, with ratings mapped to German school grades.
The grade 1 (``very good'') is the best, with the grade 5 (``inadequate'') signifying a fail.

\begin{figure}[htb]
    \centering
    \includegraphics[width=\columnwidth]{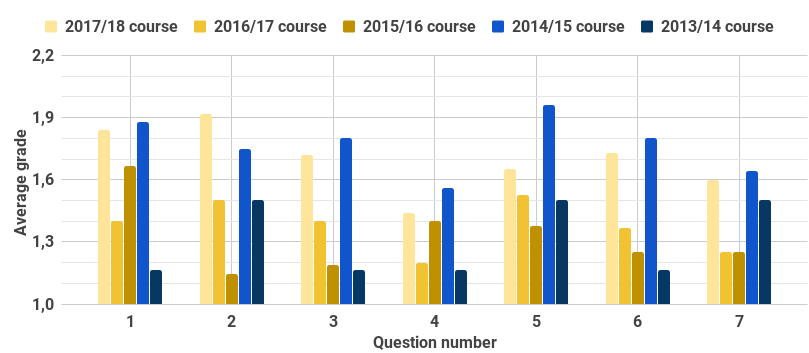}
    \caption{Mean grades given to course installments by students after the course's end. German school grades: 1 is the best grade, i.e. the lower the better. Courses in 2013/14 and 2014/15 (blue) employed solely Scrum, the others (yellow) employed both Scrum and Kanban.}
    \label{fig:evap}
\end{figure}

\textbf{Results} Average ratings for all installments of the course showed overwhelmingly positive perceptions of the course and its content, see Figure~\ref{fig:evap}.
Very few questions were answered with mean scores larger than 2 (``good'').
While this is satisfying to see as far as student satisfaction goes it also means no significant change can be detected in student satisfaction between the courses employing Kanban and those that did not.
The variance in answers is very low, as students rated aspects of the course overwhelmingly as very good (1) to good (2).
Therefore, we devised a more specific survey.

\subsection{Kanban Survey}
In the second installment of the software engineering course that used Kanban (in 2016/17), we conducted a voluntary, anonymous online survey among all students after course completion, in addition to the regular end-of-term surveys.
The first course that introduced Kanban (in 2015/16) using newly created teaching materials had received critical comments from students in oral feedback sessions, which we addressed in the following course.
The Kanban survey focused on students' perceptions of Kanban as well as the advantages and drawbacks of the introduced practices and methods.
While the survey was designed to elicit responses to the details of how Kanban was introduced in the course, it also explicitly included questions on whether and how students' workflows were adapted when changing from Scrum to Kanban processes.
These questions were designed to better understand whether the expected changes in process had actually taken place.
The survey questions related to process change are listed in Table~\ref{table:kanban_survey}.

\begin{table}[htb]
    \ra{1.2}
    \begin{tabularx}{\columnwidth}{@{}llX@{}}
        \toprule
        \textbf{\#} & \textbf{Type} & \textbf{Question}\\
        \midrule
        1 & \multicell{5-point\\scale} & Was the Kanban week at project end more useful and productive then a last week of Scrum? \\
        2 & \multicell{5-point\\scale} & Did you have to adapt your workflow for the Kanban week? \\
        3 & free text & What were the biggest advantages and disadvantages of using Kanban in your team? \\
        4 & \multicell{multiple\\choice} & How did user stories change from using Scrum to Kanban? \\
        5 & \multicell{5-point\\scale} & Would you recommend using Kanban to the participants of next year's course? \\
        \bottomrule
    \end{tabularx}
    \vspace{0.1cm}
    \caption{Questions related to Kanban adoption of the anonymous online student survey performed at the end of the 2016/17 software engineering course.}
    \label{table:kanban_survey}
\end{table}

The survey consisted mainly of questions that could be answered using a 5-point Likert scale, ranging from 1 (strong no) to 5 (strong yes), with 3 being neutral.
Additionally, the survey included free text questions as well as a multiple choice question to gather more detailed insights.
It was possible to submit the survey with missing answers.

\textbf{Results}
Overall, 18 students, 17 men and 1 woman, answered the questionnaire.
All questions featuring the Likert scale were answered by all participants.
Table~\ref{table:analysis} contains a summary of the collected answers.
\begin{table}[tb]
    \ra{1.2}
    \begin{tabularx}{\columnwidth}{@{}lXrrrrr@{}}
        \toprule
        \textbf{\#} & \textbf{Question Topic} & \textbf{Mean} & \textbf{\multicell{Std.\\Dev.}} & \textbf{\multicell{10\% Trim.\\Mean}} & \textbf{\multicell{Median}} & \textbf{Range} \\
        \midrule
        1 & Kanban week preferred over another Scrum week? & 4.08 & 1.38 & 4.30 & 5.00 & 4.00 \\
        2 & Was the workflow adapted? & 3.83 & 1.11 & 4.00 & 4.00 & 4.00 \\
        5 & Recommended for next year? & 4.33 & 0.98 & 4.50 & 5.00 & 3.00 \\
        \bottomrule
    \end{tabularx}
    \vspace{0.1cm}
    \caption{Summarized answers of participants to the 5-point Likert scale questions of the survey. Answer possibilities: 1~(strong~no), 2 (no), 3 (neutral), 4 (yes) 5 (strong yes).}
    \label{table:analysis}
\end{table}
\begin{figure}[htb]
\centering
    \includegraphics[width=0.8\columnwidth]{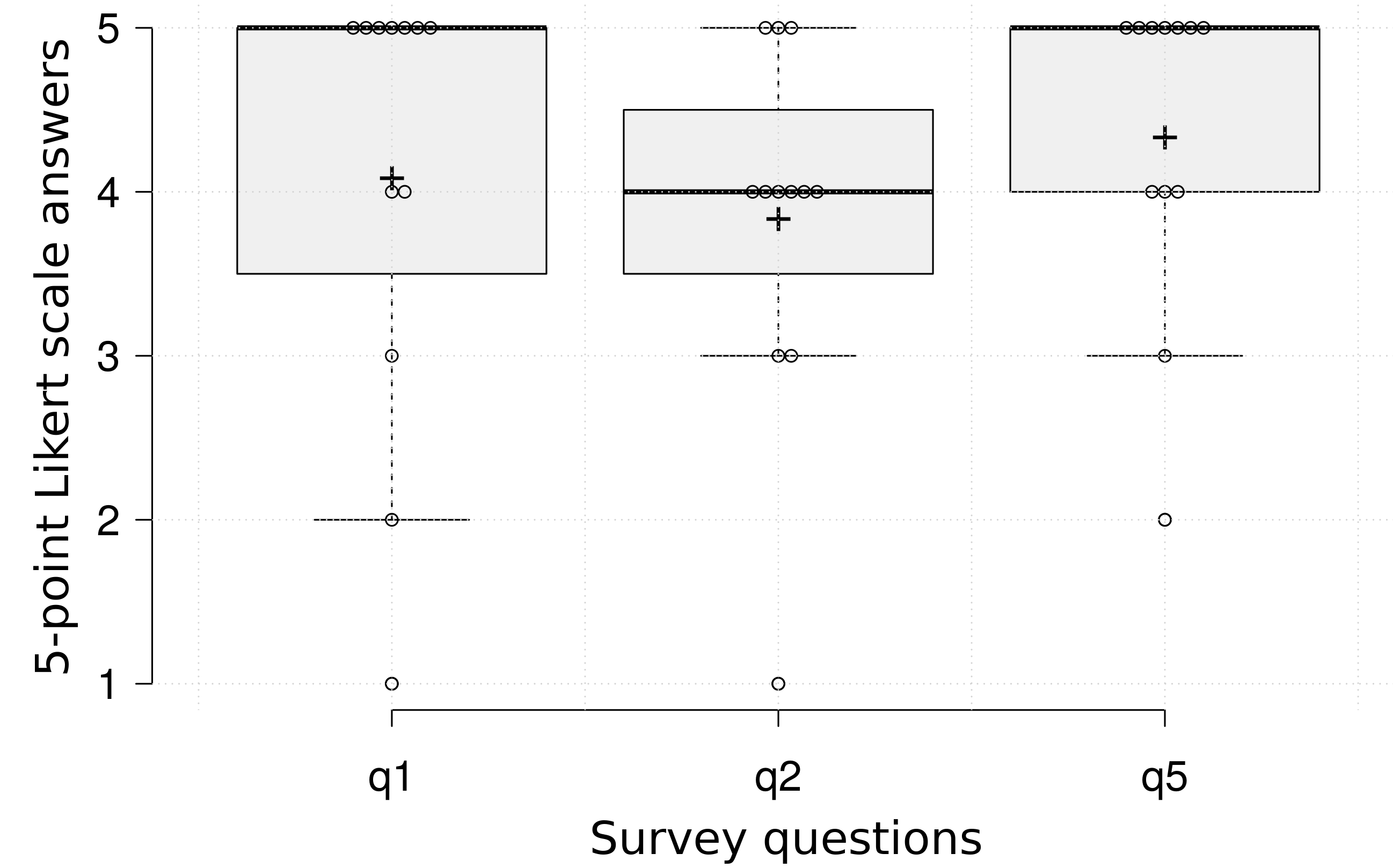}
    \caption{Summary of answers to Likert-scale questions 1, 2 and~5 as a box plot. Center lines show the medians, box limits indicate the 25th and 75th percentiles. Whiskers extend 1.5 times the interquartile range from the 25th and 75th percentiles, outliers are represented by dots. Crosses represent sample means, data points are plotted as open circles. N = 12.}
    \label{fig:boxplot}
\end{figure}
Concerning the change of Scrum to Kanban methods (question 2), students on average stated that they had adapted their workflows, see Figure~\ref{fig:boxplot}.
The high mean value (4.08), as well as the median of 5 (highest agreement), point to students having adapted their workflow in a reflected manner.
The overall positive student attitude towards group software development methodologies was also reflected in the answers to questions 1 and 6, regarding the preference of Kanban over Scrum for the last iteration as well as recommending the course for next year's students.
Surveys participants indicated that they would strongly recommend the usage of Kanban to the next cohort of students of the software engineering course (question 6), indicating that, even though this question is not a measure of learning success, applying Kanban was most likely at least fun.

The free text answers to question 3, see Figure~\ref{table:kanban_survey}, regarding the (dis)advantages of Kanban, were manually labeled with the mentioned topics.
The list of topics was refined repeatedly after evaluating every question.
Survey participants identified the following topics as advantages of Kanban (N=11):
\begin{itemize}
	\item Efficiency (7 mentions)
	\item Autonomy (4 mentions)
\end{itemize}
Three other other topics were mentioned only twice or fewer times.
As concepts, efficiency and autonomy are closely related to Lean Software's guiding principles of \emph{Eliminate Waste} and \emph{Empowering the Team}, respectively~\cite{poppendieck2003lean}.
As Kanban is heavily inspired by these ideas, it is reassuring to see that these ideas transferred.

Regarding the disadvantages of Kanban usage, students mentioned the following topics (N=9):
\begin{itemize}
	\item Only worked on small user stories (3)
	\item Uneven task distribution (2)
\end{itemize}
Another six disadvantages only received single mentions.

Solely working on small user stories, i.e. work items in an agile process, may be a consequence of team member autonomy.
Developers may choose to work on small items that can be moved through the columns of the Kanban board quickly, instead of picking larger, more time-consuming tasks to work on.
Tackling uneven task distribution between team members is an ongoing challenge in educational settings and especially in self-organizing teams of students.
It can be seen as a negative consequence of the autonomy identified by survey participants.
Developers are free to handle their workload, with some developers choosing to do more and others choosing to work on fewer items.

User stories are one of the core means of communication, both in Kanban and Scrum, between the Product Owner, who receives input from stakeholders, and developers~\cite{Rees2002}.
As such, we included a question on the perceived change of user stories when switching from Scrum to Kanban (question 4).
In order to make answering easier, this question was a multiple choice question, that provided a range of answer possibilities of which any number could be chosen.
The choices, as well as the summarized answers of survey participants, are shown in Table~\ref{table:q4}.

\begin{table}[tb]
    \ra{1.2}
    \centering
    \begin{tabular}{@{}lllll@{}}
        \toprule
        \textbf{Topic} & \multicolumn{4}{c}{\textbf{Answer choice and count}} \\
        \midrule
        User story focus & bug-oriented & 11 & feature-oriented & 0 \\
        User story length & Shorter & 11 & Longer & 0 \\
        Requirements & More detailed & 8 & More general & 0 \\
        Interaction with PO & More & 3 & Less & 0 \\
        Prioritization of stories & Better & 3 & Worse & 2 \\
        \bottomrule
    \end{tabular}
    \vspace{0.1cm}
    \caption{Answers of survey participants to question \#4, regarding attributes of user stories when changing from Scrum to Kanban processes. N=12.}
    \label{table:q4}
\end{table}

Students classified the user stories that were written by Product Owners and developers during the Kanban iteration as shorter and more bug-oriented than in the previous Scrum iterations.
While an influx of small fixes to a software product is expected shortly before the final deadline, a time in which Kanban was used, smaller user stories can also help move tickets through the Kanban board more quickly.
This reduced the \emph{cycle time}, the time from when work begins on an item until it is ready for delivery~\cite{Polk2011}.

However, students also answered that the requirements, i.e. part of the  \emph{acceptance criteria}~\cite{Silva2017} within user stories, had gotten more detailed during Kanban usage.
This allows work on a user story to be started without having to clarify open questions beforehand and can help efficiency by decreasing the cycle time.
Small user stories that contain enough detail to be immediately implementable are ideal for usage in the Kanban process~\cite{Nikitina2011}.

\subsection{Discussion}
Both surveys, the more general end-of-term survey as well as the more specialized survey on Kanban, revealed positive attitudes towards our approach of teaching agile processes in a hands-on fashion as well as the shift from Scrum to Kanban at the end of the project (\emph{RQ\ref{r2}}).
Students indicated that they would recommend using Kanban to participants of the following year's course.
These results are in line with related, similar studies~\cite{Mahnic2015}.
In particular, Melnik et al. state that students, in general, were ``very enthusiastic about core agile practices'' and accepted and liked them~\cite{Melnik2005}.
The authors also point out that this observation held for a broad range of students, regardless of educational program, age or industry experience.
Furthermore, whether the educational setting that student teams worked in during their project work was more or less controlled, did not have a significant influence on student satisfaction~\cite{Matthies2016c}.
While the findings of this and similar studies are encouraging for educators teaching software project courses, they also pose challenges.
If students are generally content in agile project courses, simply due to the course setting and the fact that teamwork and building software together is fulfilling, how can improvements to curriculum design and changes in student behavior be evaluated?

Student opinions and attitudes towards course contents are an important part of any assessment plan.
However, evaluations of teaching methods should primarily rely on the assessment of learning outcomes~\cite{Felder2009}, which surveys only partly capture as they are geared towards collecting perceptions and attitudes.
Furthermore, if not every survey participant answered the more reflective, time-consuming free text answers, data on the learning outcomes of these participants is missing completely.
To help tackle these challenges, the outcomes and artifacts produced during the process transition from Scrum to Kanban can be analyzed to gain additional insights into development teams.

\section{Development Artifact Analysis}
\label{sec:artifact}
While surveys are excellent tools for capturing the attitudes of participants, they do not allow insights into whether the perceived change in workflow or in user story quality actually took place during the project or how severe the change was.
In order to provide another dimension of analysis based on real project data, we evaluated the software development artifacts produced by course participants.
In particular, we compared students' development artifacts of the last five installments of our software engineering course, the earliest two of which did not include Kanban and the three most recent ones that did.

\subsection{Data Collection}
The development artifacts we collected from course repositories included commits into the version control system \emph{git}\footnote{\url{https://git-scm.com/}}, containing code changes, timestamps and commit messages describing the change, as well as tickets, acting as user stories, in an issue tracker.
Both of these data sources were collected from the collaboration and hosting service GitHub, where all projects were hosted.
GitHub features extensive application programming interfaces (APIs)~\footnote{\url{https://developer.github.com/v3/}} that allow programmatically extracting the data stored by the service.

For every repository of the last five course iterations, user stories/issues and commits from the last seven days of project work were collected.
This is the time frame that Kanban was employed in the more recent course iterations.

\subsubsection{Contributors}
First, all unique contributors of a project in the given time frame were identified.
This allows information extracted from different courses to be normalized by contributor count, as different course installments featured differing participant amounts.
This step was followed by manual deduplication of users, in order to merge accounts where students had used multiple accounts or email addresses, e.g. university or private accounts, to work on projects.

\subsubsection{Issues / User Stories}
Issues were only included in the analysis if they were closed in the study time frame and issues that represented GitHub pull requests\footnote{\url{https://help.github.com/articles/about-pull-requests/}} and not user stories were excluded.
For every issue, the number of comments, as well as events\footnote{\url{https://developer.github.com/v3/issues/events/}}, interactions except commenting, such as assigning developers or labels, were collected.
We furthermore annotated each issue with whether the user who had opened the issue was the same as the one who closed it.

\subsubsection{Commits}
In order to allow more rapid analysis, the git repositories of all projects under study where copied, i.e. \emph{cloned}, to a local copy.
Using the git command line, specifically the \emph{git log} sub command\footnote{\url{https://git-scm.com/docs/git-log}}, statistics on the commits where collected.
Attention was paid not to include merge commits, i.e. commits that introduce no new functionality and to take advantage of the deduplicated list of contributors.
All statistics were collected as means per contributor.
For every list of commits of a project repository, the mean commits, touched files, last-minute commits, mean line changes and the number of unique issues referenced, were saved.
Last-minute commits refer to those commits made within a day of project end.
Furthermore, we parsed the commit messages of commits and identified whether they referenced an issue in the issue tracker by number in the form ''fixed issue \#123``.

Using this data, both the assumptions on student behavior, i.e. educators hypothesis of how artifacts would change from using Scrum to Kanban, as well as the accurateness of student perceptions could be tested.

\subsection{Discussion}
The collected data shows that the length of user stories did not significantly differ from when Kanban or Scrum was used in the last iteration of the course, see Table~\ref{table:issue_length}.
\begin{table}[!htbp]
    \centering
    \ra{1.2}
    \begin{tabular}{@{}lllllll@{}}
        \toprule
        & \multicolumn{3}{c}{\textbf{Issue body length}} & \multicolumn{3}{c}{\textbf{Issue title length}} \\
        \cmidrule(l){2-4} \cmidrule(l){5-7}
        \textbf{\multicell{Course year}} & \textbf{Mean} & \textbf{Stdev} & \textbf{Median} & \textbf{Mean} & \textbf{Stdev} & \textbf{Median} \\ \midrule
2013/14 & 274.8 & 295.2 & 169.0 & 35.3 & 15.3 & 32.0 \\
2014/15 & 420.8 & 327.3 & 361.0 & 50.7 & 15.5 & 50.0 \\
2015/16\textbf{*} & 360.9 & 339.4 & 253.0 & 36.9 & 16.9 & 32.5 \\
2016/17\textbf{*} & 505.5 & 556.9 & 378.0 & 36.9 & 16.7 & 35.0 \\
2017/18\textbf{*} & 579.8 & 393.3 & 526.0 & 35.9 & 14.3 & 37.5 \\
    \bottomrule
    \end{tabular}
    \vspace{0.1cm}
    \caption{Issue body and title length of issues for the last week of projects. Courses marked with * employed Kanban.}
    \label{table:issue_length}
\end{table}

This differs from the reported perceptions of students in the survey performed in the 2016/17 course installment.
There, students reported that user stories were perceived to be shorter when using Kanban when compared to Scrum.
While these two measures are not necessarily directly comparable, further study into the content differences between user stories in Kanban and Scrum is required.
However, the analysis was able to uncover this discrepancy and provides a starting point for further investigation.

Most other measures calculated from commits, such as the mean amount of touched files, did not differ significantly between the two processes in different course years, see Table~\ref{table:commits}.
\begin{table}[!htbp]
    \centering
    \ra{1.2}
    \begin{tabular}{@{}lSSSSS@{}}
        \toprule
        \textbf{\multicell{Course\\year}} & \textbf{\multicell{Commit\\amount}} & \textbf{\multicell{Touched\\files}} & \textbf{\multicell{Last-\\minute\\commits}} & \textbf{\multicell{Line\\changes\\per commit}} & \textbf{\multicell{Unique\\issues\\referenced}}\\
        \midrule
        2013/14             & 12.1  & 13.3	& 1.4	& 590.5	& 2.8 \\
        2014/15             & 7.2	& 6.9	& 1.2	& 408.0	& 1.0\\
        2015/16\textbf{*}   & 6.1	& 8.0	& 8.1	& 466.0	& 0.1\\
        2016/17\textbf{*}   & 3.4	& 5.4	& 2.2	& 163.5	& 2.2\\
        2017/18\textbf{*}   & 8.6	& 5.3	& 1.7	& 195.0	& 1.5\\
        \bottomrule
    \end{tabular}
    \vspace{0.1cm}
    \caption{Comparison of commit attributes for the last week of projects. Courses marked with * employed Kanban. All values stated normalized by course participant count.}
    \label{table:commits}
\end{table}

However, it is encouraging to see that the amount of \emph{last-minute commits}, i.e. commits made close to the end of the iteration~\cite{Matthies2016} tended to be higher in the course installments in which Kanban was employed.
Scrum's iteration plan, the Sprint Backlog, contains all the work items a team intends to address in a sprint, i.e. the amount of work that can be performed by a team in an iteration.
Ideally, these items are worked on in a continuous manner, so that towards the end of the sprint the last user story is finished~\cite{Schwaber2017}.
In this manner, work intensity and commit frequency should be uniformly distributed during an iteration.
In contrast, Kanban does not explicitly call for iteration planning and so work is more likely to be assigned more dynamically: new work items can easily be added to the work queue, especially towards the end of the project, when the deadline approaches.

The more dynamic nature of Kanban is also reflected in the fact that the mean line change per commit, as well as the mean number of touched files, were smaller in the last two years of the course, see Table~\ref{table:commits}, when educators had already gathered some experience teaching the new methodology.
This is in line with the Kanban survey results, where students stated that their user stories when employing Kanban were more bug-oriented than feature-oriented.
Fixing a bug usually requires changing fewer lines touching fewer files than implementing an entirely new feature.
Furthermore, bugs are usually noticed during regular development activities or during testing and can easily be added to a Kanban board~\cite{Ikonen2011}, whereas they might only end up in the next Scrum sprint~\cite{Kniberg2007}.

Interactions with user stories, i.e. issues on GitHub, did not differ significantly between those courses that employed kanban and those that used Scrum, see Table~\ref{table:tickets}.
\begin{table}[!htbp]
    \centering
    \ra{1.2}
    \begin{tabular}{@{}lSSSl@{}}
        \toprule
        & \multicolumn{3}{c}{\textbf{Mean per contributor}} & \multirow{2}{*}{\textbf{\multicell{\% issues\\opened \& closed\\by same person}}}\\
        \cmidrule(l){2-4}
        \textbf{\multicell{Course year}} & \textbf{\multicell{Issue\\amount}} & \textbf{\multicell{Issue\\events}} & \textbf{\multicell{Issue\\comments}} &  \\ \midrule
2013/14 & 4.9 & 27.5 & 17.1 & 43 \\
2014/15 & 2.6 & 47.8 & 4.1 & 58 \\
2015/16\textbf{*} & 3.9 & 35.5 & 4.1 & 20 \\
2016/17\textbf{*} & 2.8 & 64.4 & 6.4 & 46 \\
2017/18\textbf{*} & 2.1 & 68.9 & 4.8 & 65 \\
    \bottomrule
    \end{tabular}
    \vspace{0.1cm}
    \caption{Comparison of issues and their attributes for the last week of projects. Courses marked with * employed Kanban.}
    \label{table:tickets}
\end{table}

Normalized by participant count, similar mean numbers of issues were closed in the studied time frame and a similar amount of comments were attached to issues.
However, the two most recent courses using Kanban showed higher mean amounts of non-comment events, such as labeling or assignments, a sign that the issue tracker was used more heavily.
While Scrum explicitly calls for a role that mainly writes and prioritizes user stories, the Product Owner~\cite{Schwaber2017}, Kanban does not.
We had thus hypothesized that introducing Kanban would result in higher engagement by the entire team with the list of outstanding work items, instead of teams mostly relying on the PO to maintain it.
However, the percentage of issues opened and closed by the same person, see the last column of Table~\ref{table:tickets}, was surprisingly low even in the Scrum courses, with only 43\% and 58\% in course installments 2013/14 and 2014/15.
These numbers did not differ significantly for the Kanban courses.
While these results do not support our original hypothesis on team engagement with user stories, they represent opportunities for future work on how agile student teams interact with user stories and work item backlogs in collaboration with a Product Owner role.

By analyzing software development data created by students, which is already produced during regular development activities, we were able to uncover areas where some of our assumptions on student behavior regarding the adoption of different agile software development methodologies were confirmed and some were refuted.
These areas will serve as a basis for future improvements to the course.

\section{Related Work}
\label{sec:related_work}
A variety of specialized data sources have been analyzed in previous work in order to gain insights into student behaviors in computer science courses.

Wilson and Shrock~\cite{Wilson2001} employed a survey to determine factors that promote success in an introductory computer science course.
They examined twelve factors of which comfort level, math, and attribution to luck for success/failure were the most important for predicting student scores.
Similarly, Bennedsen and Caspersen~\cite{Bennedsen2008} attempted to help improve students’ learning premises by collecting data on the emotional and social factors of students in a survey.
They collected data on the factors of perfectionism, self-esteem, coping tactics, affective states, and optimism.

While surveys can be used to collect data on arbitrary factors, depending on which questions are used, entirely different ways of collecting student data have also been explored.
Keen and Etzkorn~\cite{Keen2009} analyzed the complexity of teacher's lecture notes, the \emph{buzzword density}, i.e. the amount of Computer Science domain-specific words divided by the total number of words in the lecture, to predict grades.
Fire et al.~\cite{Fire2012} built a graph of social interaction between students using homework assignment data and course website logs. 
The authors then studied this data structure to reveal the impact of cooperation among students on their success.
The model showed a high correlation between a student's grade and that of their closest friend.
Ashenafi et al. used data created by students during course activities, namely the results of several semi-automated peer-assessment tasks throughout the semester, to build a linear regression model for predicting final grades~\cite{Ashenafi2015}.

Performance prediction has more recently also been applied in MOOCs, which feature an abundance of digital data on student behavior.
Jian et al.~\cite{Jiang2014} used students' activity in discussion forums together with their performance on peer-assessment tasks and assignments in the first week of the course to predict whether students will complete the course.
Brinton et al.~\cite{Brinton2015} collected behavior data of students watching the educational videos of a MOOC.
They recorded interactions with the video player, e.g. pause, rate change or seek in order to predict whether participants will correctly answer questions on a video's content in the first attempt.
Teusner et al.~\cite{Teusner2017a} compiled events of users interacting with a MOOC platform in the form of standardized events~\cite{DelBlanco2013}.
From these event streams 17 combined metrics, such as session duration, forum, video player and download activity, as well item discovery are computed.
The authors state that using the collected data, \emph{informed actions} can be taken to improve learning outcomes.

\section{Conclusion}
\label{sec:conclusion}
University courses need to be continuously adapted to changes in requirements, such as industry shifts, technology advancements or altered student expectations.
Feedback to educators on curriculum design usually takes the form of end-of-term surveys or questionnaires on specific course aspects.
However, techniques from the field of Educational Data Mining can be employed to gain insights into student teams and their reactions to changes in curricula, that go beyond those possible with surveys.
This paper presents our approach to analyze development artifacts to achieve this goal.
We analyzed the software development artifacts of student teams from five university undergraduate software development courses, which teach different agile development methodologies.
Surveys with participants revealed positive attitudes towards the course and changing the employed development methodology during the course from Scrum to Kanban.
However, surveys were not able to ascertain the degree to which students had adapted their workflows accurately.
Therefore, we propose an approach of analyzing the software development data created by students during regular project work activities, specifically user stories and commits to a version control system, as an additional dimension of analysis.
While this data serves a primary purpose in communication between students it also represents information that can be collected and analyzed by educators to generate insights into student behavior during project work.

\balance

\bibliographystyle{IEEEtran}
\bibliography{library}

\end{document}